\documentclass[journal=nalefd,manuscript=article]{achemso}

\usepackage[version=3]{mhchem} 
\usepackage[T1]{fontenc}       

\author{Abhinav Kumar}
\affiliation[]
{Department of Applied Physics, The Hong Kong Polytechnic University, Hong Kong SAR}
\author{Alejandro H. Strachan}
\affiliation[]
{School of Materials Engineering and Birck Nanotechnology Center, Purdue University, West Lafayette, Indiana 47906, USA}
\author{Nicolas Onofrio}
\email{nicolas.onofrio@polyu.edu.hk}
\phone{(+852) 2766 5680}
\fax{(+852) 2333 7629}
\affiliation[]
{Department of Applied Physics, The Hong Kong Polytechnic University, Hong Kong SAR}

\title[]
  {Prediction of low energy phase transition in metal doped MoTe$_2$ from first principle calculations}


\begin{document}






\begin{abstract} 
Metal-insulator transitions in two dimensional materials represent a great opportunity for fast, low energy and ultra-dense switching devices.
Due to the small energy difference between its semimetallic and semiconducting crystal phases, phase transition in MoTe$_2$ can occur
with an unprecedented small amount of external perturbations.
In this work, we used density functional theory to predict critical strain and electrostatic voltage required to control the phase transition
of 3d and 4d metal doped MoTe$_2$.
We found that small doping contents dramatically affect the relative energies of MoTe$_2$ crystal phases and can largely reduced
the energy input to trigger the transition, compared to pristine case.
Moreover, the kinetics corresponding to the phase transition in the proposed doped materials are several order of magnitude faster than in MoTe$_2$.
For example, we predict 6.3 \% Mn doped MoTe$_2$ to switch phase under 1.19 V gate voltage in less than 1 $\mu$s with an input energy of  0.048 aJ/nm$^3$.
Due to the presence of dopant, the controlled change of phase is often complemented with a change in magnetic moment leading
to multi-functional phase transition.
\end{abstract}

Materials exhibiting metal-insulator transition, which is characterized by a sudden change in resistance and transmittance near transition,
have shown a vast range of applications including rewritable optical storage media\cite{mem-2} and switches\cite{op-1}, thermochromic windows\cite{window-1},
laser protection\cite{laser-1}, energy harvesting systems\cite{ene-harv} and, memory devices\cite{mem-1}.
For example, transition metal oxides such as VO$_2$\cite{IEEE-TMO} and, chalcogenide glasses\cite{hudgens2004overview} (e.g. Ge$_2$Sb$_2$Te$_5$) have recently
driven great attention for application as resistive switching devices due to their controllable phase transition.
However, significant temporal and spatial variations of the switching voltages and resistance states, poor retention time and, thermal disturbances
remain major issues to address\cite{IEEE-TMO,lacaita2006phase}.
Moreover, the downscaling limit imposed by current materials estimated to reach approximately 10 nm and their high power are far from next-generation
nanoelectronics requirements.

Group VI transition metal dichalcogenides (TMDs) represent a special class of fast-growing two dimensional (2D) materials\cite{chhowalla2013chemistry}
with various electronic properties depending on the number of layer, their composition and phase.
Single-layer TMDs have been used to make atomically thin field effect transistors\cite{radisavljevic2011single} and, nanophotonic\cite{photo-trans,ling2015prospect}
and optoelectronic devices\cite{optoelec,jariwala2014emerging}.
Phase transition in TMDs has been studied for years mostly by thermal and chemical means\cite{keum2015,Kappera2014}.
Unlike their bulk form, few layered molybdenum dichalcogenides have direct bandgaps in their ground state H-phase, which decreases
from sulfur to tellurium\cite{band-gap-1, band-gap-2}.
By contrast, tungsten ditelluride exhibits a semi-metallic ground state T'-phase\cite{WTe2-semi}.
A small amount of strain in the range 3 to 0.3 \% or, an electrostatic voltage of 1.8 V, have been predicted to induced phase transition
in monolayer MoTe$_2$ from H to T'-phase\cite{Reed-14}.
This semiconductor-to-metal transition was demonstrated experimentally with a reported strain of 0.2 \%, at room temperature\cite{exp-strain}
and later via an applied gate voltage of 3 V\cite{nat-exp-reed}, in good agreement with the theoretical predictions.
Although mechanical phase transitions are of interest\cite{nayak2014pressure,exp-strain}, electrostatic doping remains the most appealing pathway
for electronic device applications.
Electrostatically-driven phase transition in MoTe$_2$ has shown potential for high performance switching devices with predicted
energy consumption per unit volume 100 to 10,000 times lower than the well-studied Ge$_2$Sb$_2$Te$_5$, at T=300 K\cite{reed-npj}.
However, the reported energy barrier associated with MoTe$_2$ phase transition is 0.88 eV/f.u. corresponding to 50 s timescale, at room temperature\cite{Reed-14}.
This slow kinetic and, high voltage operation are still above standards of nowadays semiconducting industry.

Interestingly, it has been shown that alloying MoTe$_2$ with W can reduce the transition voltage\cite{nano-W-dop,santosh,Reed-16}.
Furthermore, doping has the advantage of local control which can be used in homojunction devices\cite{Cho625} where the contact and the device are made
of the same material.
Ground state H-phase MoTe$_2$ has similar bandgap as silicon and, its semi-metallic T'-phase exhibits topological properties which can be used
as dissipation-free contacts in homojunction devices.
However, beside W doped MoTe$_2$, only Re doped MoSe$_2$\cite{vidya}, Sn doped MoS$_2$\cite{Sn-dop}, and the effect of chalcogen
substitution or vacancy\cite{chalcogen-sub} have been reported to engineer the phase transition.
Despite the fact that alloying has the potential to reduce the transition voltage, and therefore the energy consumption over pristine
materials\cite{santosh,nano-W-dop}, we lack a mechanistic understanding of the effect of dopants to the phase transition.

In this paper, we used high-throughput density functional theory (DFT) calculations to explore various 3d and 4d transition metals (TM) as low
concentration dopants in MoTe$_2$.
We show that only a small amount of TM substituent can lower the phase transition energy and, further application of small external perturbations
such as strain or electrostatic doping can trigger the reversible phase transition.
For example, we predict T' to H-phase transition in 6.3 \% Mn-doped MoTe$_2$ with a small amount of strain of only 0.63 \% (no vibrations) along
the a-direction or, a voltage gating of 1.19 V.
These values are significantly smaller than for pristine MoTe$_2$ predicted to be 2.4 \%\cite{Reed-14} (along b) and 1.8 V\cite{Reed-16}.
The lower transition voltage predicted in Mn doped MoTe$_2$ compared to pristine case corresponds to an energy consumption reduced by approximately 40 \%.
Moreover, we predict the kinetics of the phase transition for some key dopants to be approximatively 10$^5$-10$^7$ times faster than in MoTe$_2$,
faster than actual Flash memory\cite{waser2016introduction}.
To generalize, we propose a simple model to estimate the critical dopant concentration at which phase transition will occur.
We further introduce multi-functional phase transition by showing the complementary semiconductor-to-metal electronic transition with magnetic state
transition as the magnetic moments in H and T'-phases are different for most of the dopants studied.

\section{Results and discussion}

\begin{figure}[h]
  \centering
  \includegraphics[width= 0.9\textwidth, angle=0]{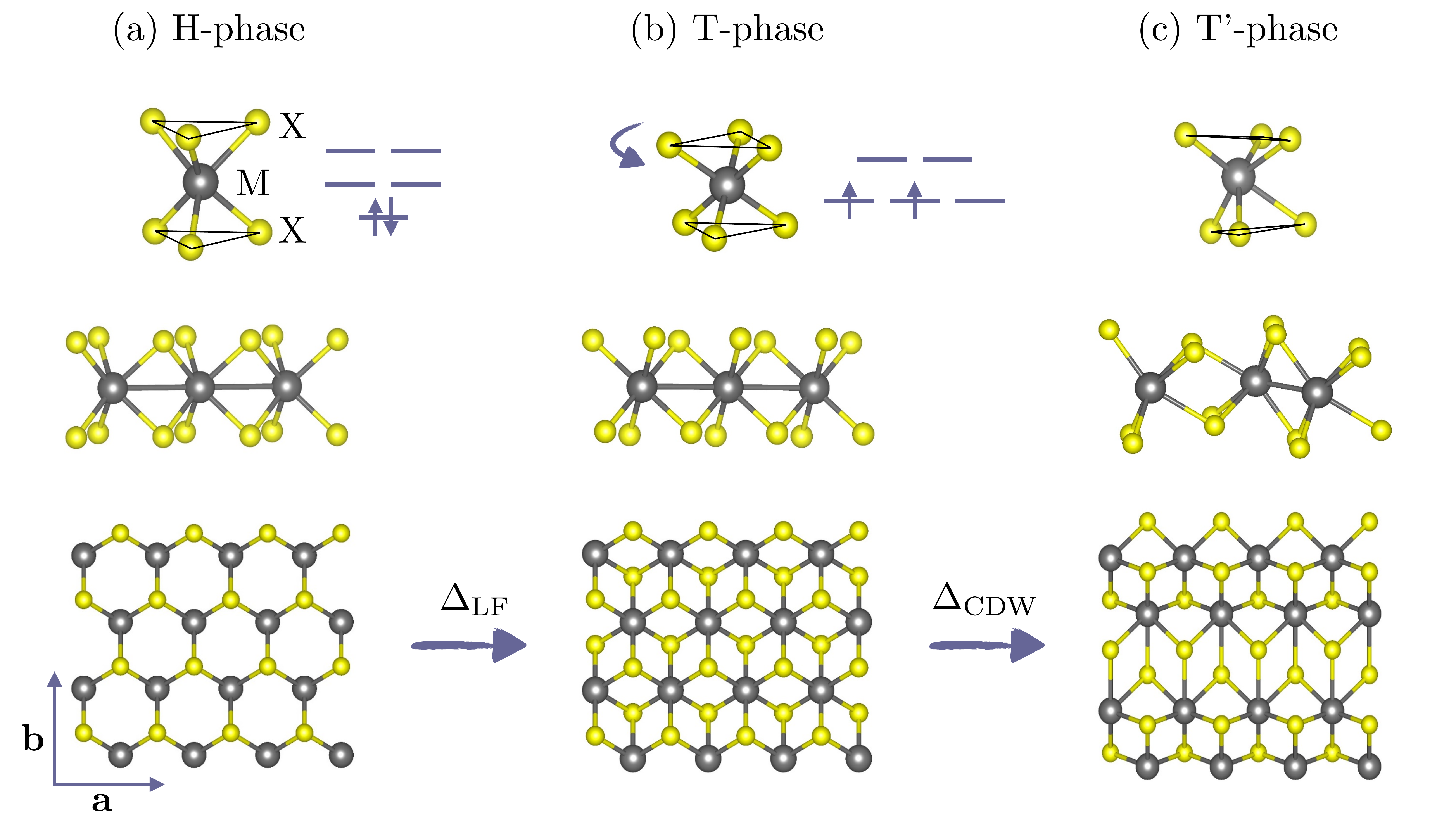}
  \caption{Crystal structures and ligand field splitting of H (a), T (b) and T'-phases of monolayer TMDs of formula MX$_2$.
The fully and partially occupied d-orbital(s) in the H and T (and T') phases leads to semiconducting and metallic characters.
The transition from H to T-phase occurs via a 60$^\circ$ rotation of the upper chalcogen layer and is characterized by a change
in the ligand field energy $\Delta E_{LF}$.
The transition from T to T'-phase is induced by Peierls-like distortions leading to charge density wave energy stabilization $\Delta E_{CDW}$.
Gray and yellow spheres represent metals and chalcogen atoms, respectively.}
  \label{fig:structure}
\end{figure}

{\bf Structural details.}
The interplay of two mechanisms determines the ground state phase of group VI TMDs known as ligand field splitting (LFS) and,
charge density wave (CDW) \cite{santosh}.
Monolayer TMDs, have general formula MX$_2$ and are composed of a layer of TM sandwiched between two layers of chalcogens ions.
In the most common H-phase, each metal ion is surrounded by 6 chalcogen ions making a trigonal prismatic environment which
splits the d-level of the metal into 1-fold and two 2-fold levels, as shown in Figure \ref{fig:structure}a.
As the lowest level is completely occupied by the two d-electrons from the metal, the H-phase of TMDs are semiconducting in nature.
The geometry of the T-phase can be visualized as a rotation of one of the chalcogen layer (top layer in Figure \ref{fig:structure}b)
about the c-direction by 60$^\circ$.
The environment of the TMs change into octahedral and the corresponding crystal field splitting now becomes 3-fold and 2-fold.
Partially filled d-levels results in a metallic state.
The energy difference between H and T-phases depends on the strength of the LFS and, is usually referred as $\Delta E_{LF}$.
The increased M-X bond lengths, when going from sulfide to telluride, reduces the LFS and hence the bandgap.
The high density of state at the Fermi level in the metallic T-phase induces a Peierls-like distortion similar to VO$_2$,
in which dimerization and a tilt of dimers have been observed.
This new phase which is the ground state for WTe$_2$ is known as distorted T-phase or T'-phase.
But unlike VO$_2$, the T'-phase of TMDs are semi-metallic in nature.
The energy difference between T and T'-phases is referred as $\Delta E_{CDW}$.
Due to the richness of energetically close crystal phases, TMDs represent perfect candidates for ultra-dense phase
change memory materials.

\begin{figure}[h]
  \centering
  \includegraphics[width= 0.9\textwidth, angle=0]{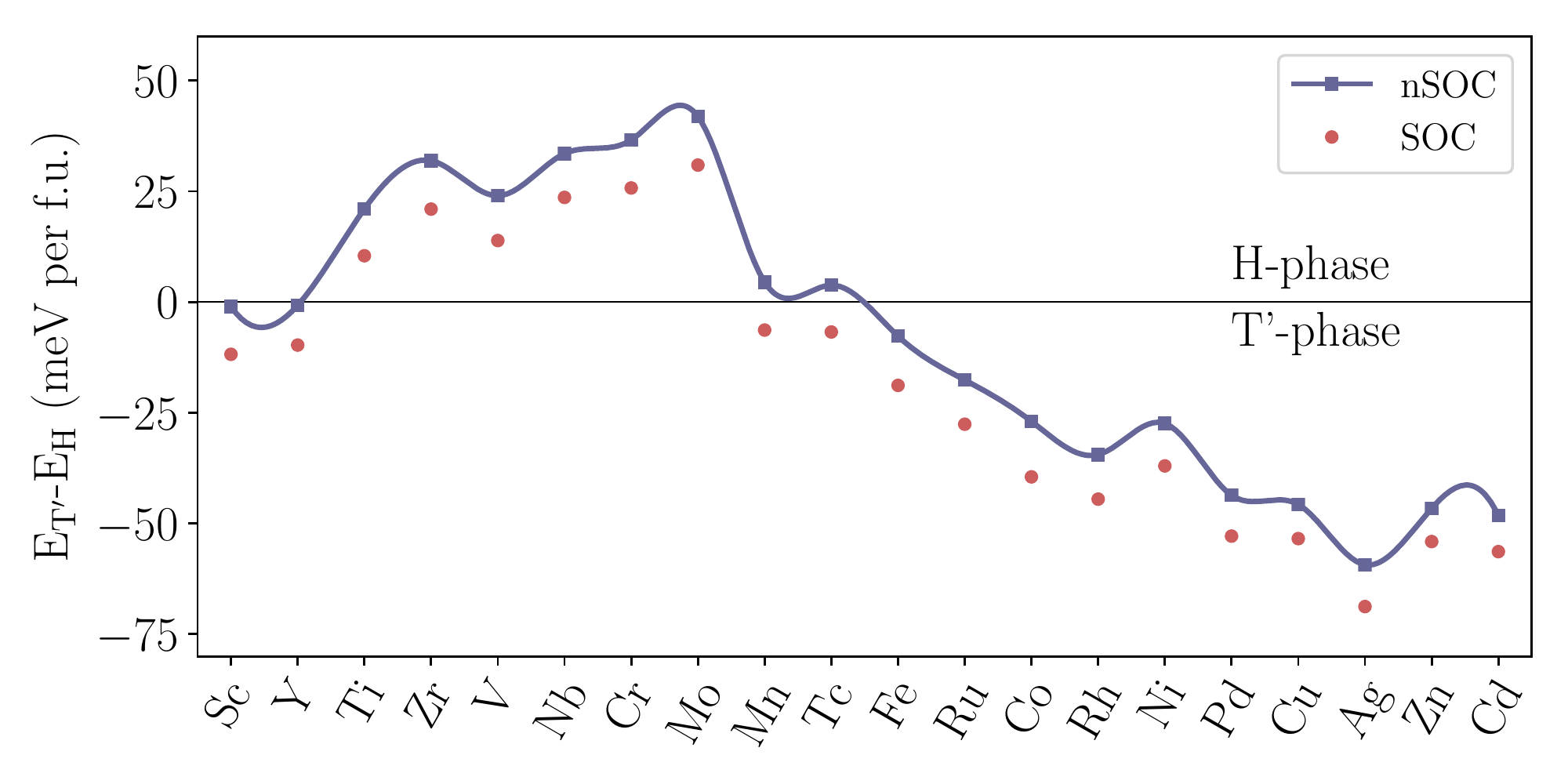}
 \caption{Energy difference between T' and H phases per formula unit of MoTe$_2$ for various (6.3 \%) dopants computed with (red circles) and
without (blue squares) spin-orbit coupling.
Each phase has been fully relaxed including ions and lattice parameters.}
  \label{fig:TM-dop}
\end{figure}

Among TMDs, MoTe$_2$ has the smallest H/T' energy difference\cite{santosh}.
Various doping types have been discussed as a means to stabilize the semi-metallic T'-phase in TMDs.
Knowing the fact that TM dopant prefers Mo substitution in all molybdenum dichalcogenides\cite{dft-nico,dft-sunega,dft-sanvito}, we have doped
3d and 4d TMs in MoTe$_2$ and, studied the effect of doping on the stability of H and T'-phase.
We started from a supercell with 16 MoTe$_2$, as shown in Figure \ref{fig:structure} and, replaced one Mo
with various TMs, corresponding to 6.3 \% of doping ($\sim$6$\times$10$^{13}$ cm$^{-2}$).
This doping concentration is routinely achieved via chemical vapor deposition for various electronic applications\cite{suh2014doping,azcatl2016covalent}.
Figure \ref{fig:TM-dop} shows the energy difference $E_{T'}-E_{H}$ per formula unit for various dopants
computed with DFT (at the PBE level) with (red) and without (blue) spin-orbit coupling (SOC).
Details on the calculations are provided in the Method section.
Maxima in the curves indicate pristine MoTe$_2$ followed by the isovalent Cr doped, at slightly lower energy difference.
The right side of a maximum, which has a steeper slope, corresponds to dopants having more electrons than the pristine case and,
the left side corresponds to dopants having less electrons.
The shape of the curve can be explained on the basis of the difference in the electronic structure of the two phases.
The extra electrons added to the H-phase must cross the bandgap to reach the available conduction band.
This costs more energy than the extra electrons added to the T'-phase which accumulate near the Fermi energy, due to its semi-metallic nature.
By contrast, removal of electrons reduces the energy of the T'-phase more than that of the H-phase but, this effect is smaller and so the slope is less steep.
The inclusion of SOC lowers the H/T' energy differences by approximately 7-10 meV for the various dopants.
The effect of SOC is critical, especially for Mn and Tc doped systems, for which its inclusion switches the ground state structure from H to T'.
Interestingly, Figure \ref{fig:TM-dop} shows that only a small amount of TM dopants (e.g. 6.3 \% Sc or Y) can dramatically affect the phase stability
of MoTe$_2$ by contrast with the large amount of 33 \% W predicted Ref.\cite{santosh}.
This can be explained by the isolvalent nature of W which induces only small LFS and contributes to the energy transition via CDW.
Only Ti, Zr, V, Nb and Cr dopants stabilize the H-phase while all the other TM tested show the T'-phase thermodynamically more stable.
TMs represent a unique group of dopants to finely tune the phase transition in MoTe$_2$ as suggested by the almost continuous spectrum
of energy differences.
We now predict the critical voltage and strain to trigger the phase transition for each doped MoTe$_2$ and, discuss the consequences.

\begin{figure}[ht]
\includegraphics[width= 0.75\textwidth, angle=0]{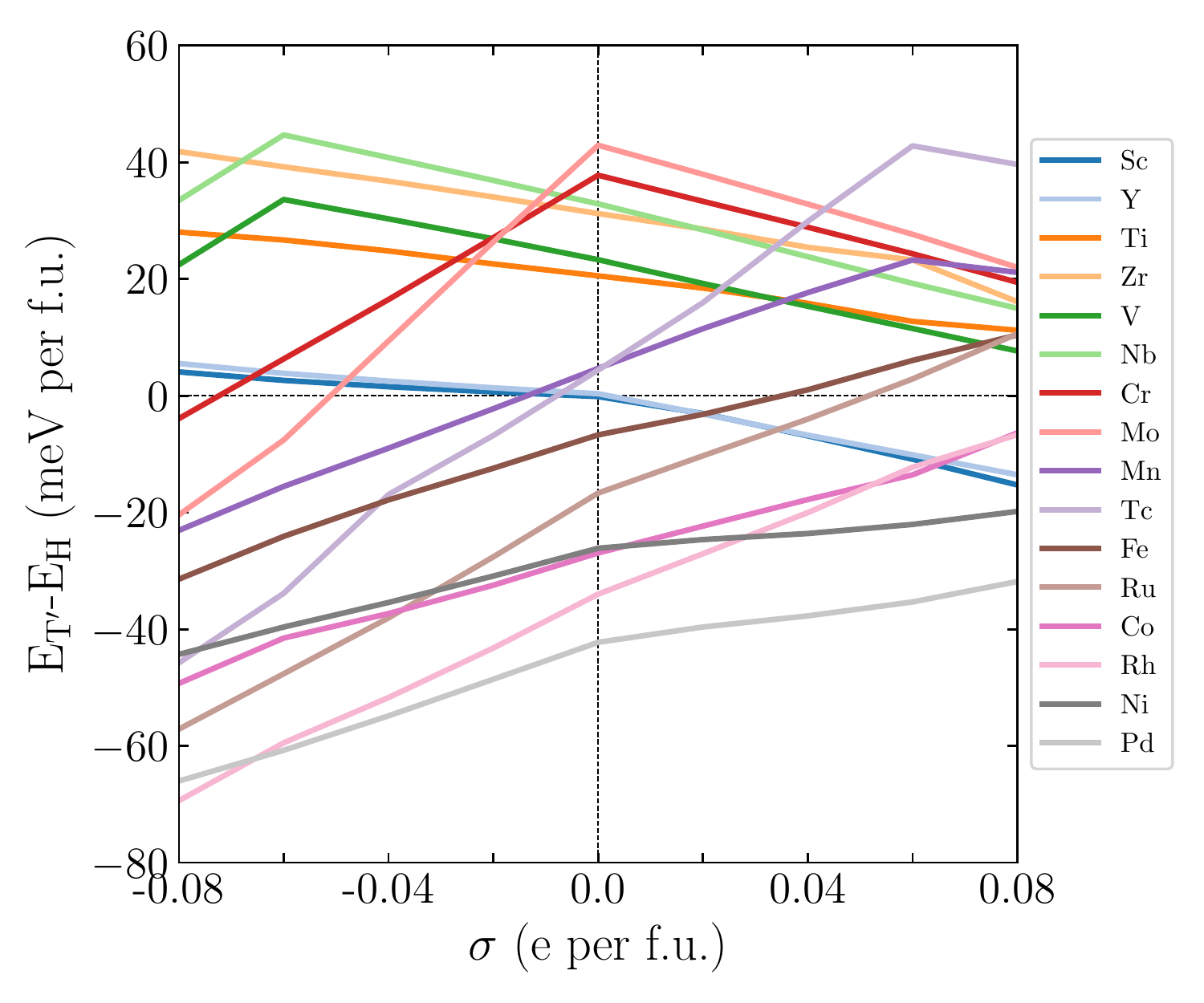}
\caption{Energy difference between T' and H phases per formula unit of MoTe$_2$ without SOC for various TM dopants (except Cu, Ag, Zn and Cd)
as a function of excess charge density per MoTe$_2$.}
\label{fig:charge}
\end{figure}

{\bf Charge doping.}
A way to dynamically control the phase transition in MoTe$_2$, which has been discussed in details Ref. \cite{Reed-16},
relies on electrostatic gating of a capacitor like structure, where the MoTe$_2$ monolayer has been deposited to one of the two plates.
We performed charge doping calculations of all TM doped systems and evaluated the constant stress phase stability of the materials.
The constant stress condition corresponds experimentally to low friction between the monolayer TMD and the substrate\cite{Reed-16}.
The number of electrons in these doped materials was varied within the range -0.08 to +0.08 e/f.u. (by step of 0.02 e/f.u.).
Section S1 of the Supplementary Materials describes how one can convert the charge density to the applied voltage.
Figure \ref{fig:charge} shows the H/T' energy difference as a function of charge density doping for all TM doped MoTe$_2$ without SOC.
As charge is added or withdrawn from the compounds, the corresponding H/T' energy difference adjusts through the relative stabilization of
each phase, similar to the mechanism described above upon doping.
Isovalent Nb and V have maxima shifted by -0.06 e/f.u. which corresponds to $\sim$1/16 e withdrawn from pristine MoTe$_2$
(i.e. one of the 16 Mo substituted).
Similarly, Mn and Tc are shifted to +0.06 e/f.u., corresponding to the addition of an electron per supercell.
Therefore, charge doping is equivalent to substitutional doping without the effect of local strain of the substituent (i.e. LFS).
Furthermore, the shape of the curves presented Figure \ref{fig:charge} are related to the detailed electronic structure of the doped materials
as suggested by the difference in slope between pristine MoTe$_2$ and the Cr-doped (isovalent) compound.

As discussed above, the inclusion of SOC is important for quantitative predictions.
To reduce the computational cost of SOC calculations, we made the assumption that the effect of SOC in charged materials is similar
to that in the neutral case.
Therefore, we added the difference between SOC and non-SOC (nSOC) energies of the neutral system for each phase and each
dopant to the corresponding nSOC energies of each charged systems.
To validate this assumption we also computed some cases including fully SOC.
A comparison of critical excess charge density (the charge density required to trigger the phase transition) for nSOC and SOC-corrected
calculations (SOC$^*$) is reported Table \ref{table:charge}.
We found critical charge densities of +0.160 and -0.051 e/f.u. for pristine MoTe$_2$ without SOC.
These values reduced to +0.120 and  -0.038 e/f.u. with SOC$^*$ correction, consistent with previous reports\cite{santosh,Reed-16} as well as
our fully SOC calculations (SOC column of Table \ref{table:charge}) corresponding to +0.112 and -0.040 e/f.u., respectively.

\begin{table}[h]
\begin{tabular}{c|ccc|ccc|}
  & \multicolumn{6}{c|}{Critical charge density (e per f.u.)} \\
  & \multicolumn{3}{c|}{$\sigma_{+}$} & \multicolumn{3}{c|}{$\sigma_{-}$ }  \\
  Dopant & nSOC & SOC$^*$ & SOC & nSOC & SOC$^*$ & SOC \\ \hline \hline
 Sc & n/a & n/a & & {\bf -0.004} & n/a &  \\
 Y  & {\bf +0.002} & n/a &  & n/a & n/a &  \\
 Ti & n/a & n/a &  & n/a & n/a &  \\
 Zr & n/a & n/a &  & n/a & n/a &  \\
 V  & n/a & {\bf +0.067} &  & n/a & n/a &  \\
 Nb & n/a & n/a &  & n/a & n/a &  \\
 Cr & n/a & n/a &  & {\bf -0.072} & {\bf -0.051} &  \\ \hline
 Mo & +0.160 & +0.120 & +0.112$^{a}$ & {\bf -0.051} & {\bf -0.038} & {\bf -0.040}$^{a}$ \\ \hline
 Mn & n/a & {\bf +0.018} & {\bf +0.019}$^a$ & {\bf -0.014} & n/a &  \\
 Tc & n/a & {\bf +0.011}  & {\bf +0.012}$^a$ & {\bf -0.008} & n/a &  \\
 Fe & {\bf +0.035} & {\bf +0.080} &  & n/a & n/a &  \\
 Ru & {\bf +0.052} & {\bf +0.079} &  & n/a & n/a &  \\
 Co & n/a & n/a &  & n/a & n/a &  \\
 Rh & n/a & n/a &  & n/a & n/a &  \\
 Ni & n/a & n/a &  & n/a & n/a &  \\
 Pd & n/a & n/a &  & n/a & n/a &  \\
 \hline
 \end{tabular}
\caption{Positive ($\sigma_+$) and negative ($\sigma_-$) critical charge density per formula unit required to switch the phase of various doped
MoTe$_2$ with (SOC) and without SOC (nSOC).
The column SOC$^*$ corresponds to an approximation of the SOC case, as described in the text.
nSOC values were extracted from Figure \ref{fig:charge} SOC from Figures S3 and S6 via linear interpolation of the intersecting curves.}
\label{table:charge}
\begin{flushleft}
$^a$ from Figures S3 and S6.
\end{flushleft}
\end{table}

Based on this, we predict that V, Mn and Tc dopants require lower positive critical charge densities to change phase compared to pristine MoTe$_2$.
Both H and T' ground state phases are observed, depending on the dopant.
Interestingly, we found that small positive critical excess charges of only +0.011 and +0.018 e/f.u. would trigger the phase transition in
6.3 \% Tc and Mn-doped MoTe$_2$, respectively, in good agreement with the SOC values of +0.012 and +0.019 e/f.u., validating further the approximation.
These small excess charges correspond to critical (constant voltage mode) voltages of 1.20 and 1.19 V, respectively and significantly lower than our predictions
for pristine MoTe$_2$ of +3.75 and -1.29 V.
The corresponding energy consumption to trigger the phase transition are 0.11, 0.048 and 0.062 aJ/nm$^3$ for pristine, Mn and Tc doped MoTe$_2$,
respectively (details in the Supplementary Materials, section S1).
Moreover, Ti, Fe and Ru-doped materials also show small critical charge densities ($<$0.1 e/f.u.) and represent additional candidates to tune the H/T' phase transition.
Isovalent Cr dopant exhibits similar electrostatic behavior than pristine case with a predicted phase transition at slightly larger negative charge density value.
All TM dopants are predicted to require larger negative charge density to trigger the transition compared to that for pristine MoTe$_2$.
It is worth to mention that these critical charge densities do not account for any thermal effect and should therefore be considered as an upper limit.
A recent discussion of such effects in MoTe$_2$ and, the corresponding effect on the energy consumption can be found Ref.\cite{reed-npj}.

We performed additional simulations to evaluate the effect of dopant concentration to the H/T' energy difference.
Figure S7 shows H/T' nSOC energy differences as a function of dopant type corresponding to 12.5, 6.3 and 4.2 \% doping concentrations.
As expected, the energies are dramatically affected by the concentration of the dopant.
This provides additional candidates exhibiting small critical charge density to trigger the phase transition.
For example, at 4.2 \% dopant, Sc, Y, Mn, Tc, Fe and Ru appear now to be in the ground state H-phase, approximately 10 meV above the T'-phase.
However, the inclusion of SOC shows that 4.2 \% Mn doped MoTe$_2$ is T'-phase ground state and, we predict a small critical charges of 0.009 e/f.u.
to switch to the H-phase.
By contrast, 4.2 \% Ti doped MoTe$_2$ is H-phase ground state and requires only 0.007 e/f.u. to switch to the T'-phase.
To generalize, we show Figures S8-S9 that the H/T' energy differences are approximately proportional to the dopant concentration, providing
a simple way to estimate the critical doping concentration.
Table S1 summarizes (nSOC) critical doping concentrations required to trigger the phase transition in various doped MoTe$_2$.
This can be easily extended to include SOC effect.

Although Figures \ref{fig:TM-dop} and S7 provide important informations related to the thermodynamic stability of each phase, they lack
kinetic informations.
We performed nudged elastic band (NEB) calculations to evaluate the energy barriers between H and T' phases of some key dopants.
The corresponding potential energy surfaces are reported in the Supplementary Materials, Figure S10.
We found activation energies of 0.54, 0.41 and 0.34 eV for 6.3 \% Tc, Mn and Sc doped MoTe$_2$, respectively.
If we consider a characteristic frequency of 10 THz\cite{Reed-14}, these barriers correspond to 101.0 $\mu$s, 0.9 $\mu$s and 40.8 ns transition
timescales respectively, 6 to 8 orders of magnitude faster than the pristine case (details in the Supplementary Materials, section S3).
These transition speeds are faster than state-of-the-art Flash non-volatile memory and approach that of DRAM\cite{waser2016introduction}.

\begin{figure}[ht]
\includegraphics[width=0.75\textwidth]{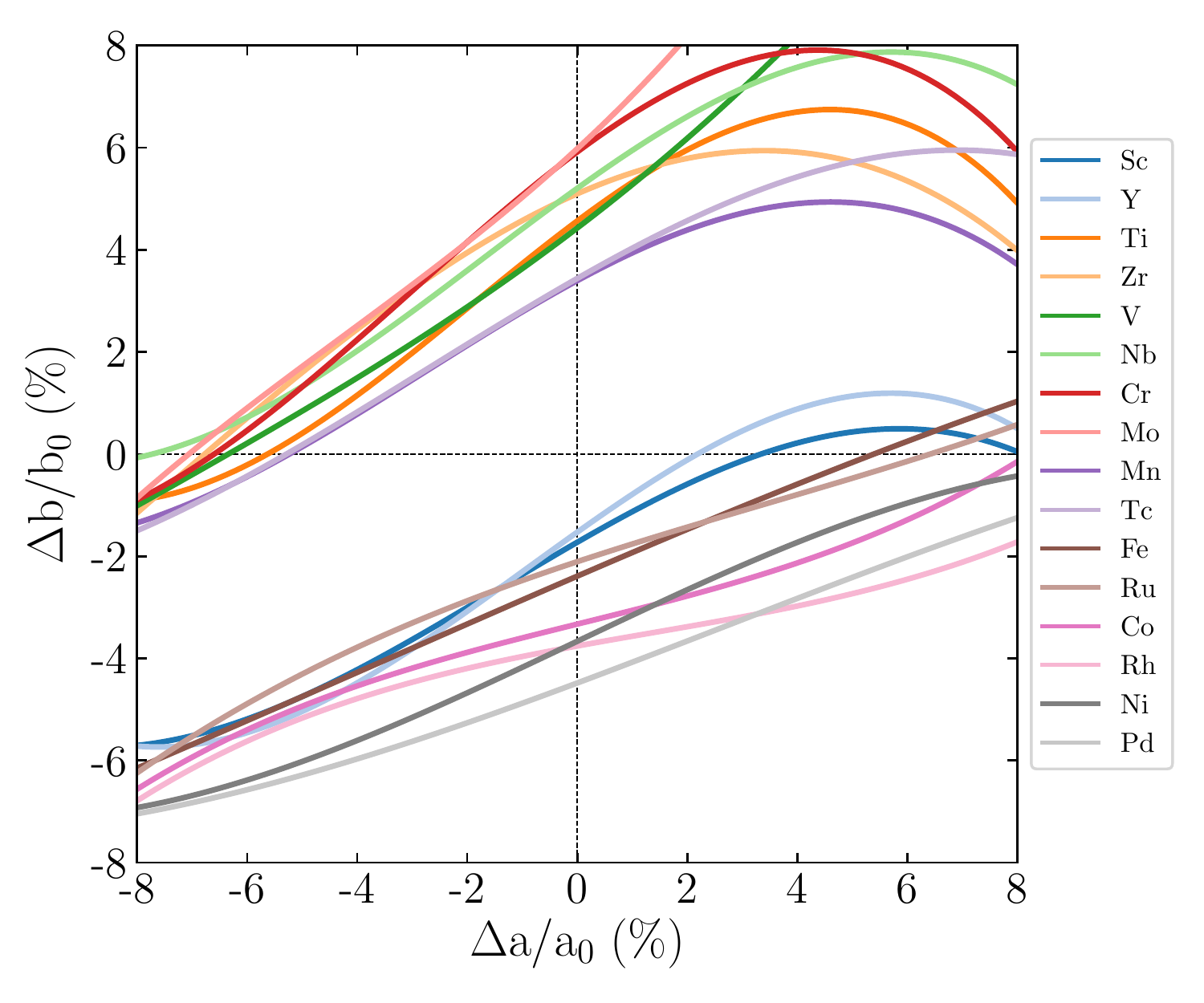}
  \caption{Intersection contours of the H and T' energy surfaces when varying the lattice constants a and b of various doped MoTe$_2$ monolayers.
The lattice constants a and b are represented as percent engineering strains, normalized over the equilibrium lattice constants of the ground state phase a$_0$ and b$_0$.
These calculations do not include spin-orbit coupling.}
 \label{fig:strain}
\end{figure}

{\bf Strain effect.}
Another way to induce phase transition in these materials, discussed Ref.\cite{Reed-14}, is based on strain.
To investigate the effect of strain on the stability of each doped MoTe$_2$ compound, we calculated the energy of the corresponding H and T' phases
on a 5$\times$5 grid of strain ranging between -8 to +8 \%, around the ground state relaxed cell.
The corresponding energies were interpolated and the intersection of the H and T' surfaces, normalized with respect to zero strain lattice parameter of the
ground state doped system, are reported Figure \ref{fig:strain}.
All doped material with H-phase ground state, cluster together in the upper quadrants of the figure.
The intersection contours corresponding to these H-phase ground state materials cut the positive region of the b-axis and, that of the negative a-axis.
Similarly, the intersection curves of dopants with T'-phase ground state cluster together, lie in the lower quadrants of the figure and, cut the negative
values of the b-axis and positive values of the a-axis.
This suggests that a tensile strain (positive) along $b$ or a compressive strain (negative) along $a$ directions is required to switch phases for
H-phase ground state compounds while a tensile strain along $a$ or a compressive strain along $b$ directions is needed to trigger transition of the T'-phase ground state systems.

We evaluated the uniaxial load required along $a$ and $b$ directions to switch phase in the 6.3 \% doped MoTe$_2$ monolayers.
The procedure to extract uniaxial loads is detailed section S4 of the Supplementary Materials.
We also evaluated the effect of SOC to the critical uniaxial load by considering a constant correction from the unstrained crystal phase,
in the same spirit as the correction on critical excess charge described before.
Details are provided in the Method section.
The critical values of strain along a and b directions are presented in Table \ref{table:uni-load}.
Keeping in mind that the experimental realization of a negative load can be challenging, we tried to identify cases with small positive load.
We found a critical load value of 2.49 \% for pristine MoTe$_2$ along b-axis without SOC which matches well with a previous study finding
2.40 \%\cite{Reed-14}.
This value reduces to 1.88 \% by including the approximated SOC correction (SOC$^*$) in perfect agreement with the fully SOC calculations
we performed leading to the critical strain of 1.84 \%.
Our simulations predict small critical uniaxial strains of -0.35 and -0.55 \% along $a$ and 0.30 and 0.46 \% along $b$ to trigger the phase transition of
Ti and V doped MoTe$_2$, respectively.
The T'-phase ground state Tc and Mn compounds exhibit phase transition upon 0.56 and 0.53 \% along $a$ and -0.25 and -0.34 \% along $b$, respectively.
This was validated with SOC calculations and we found between 0.03-0.47 \% error in the SOC$^*$ predictions.
These strain values are significantly lower than for pristine MoTe$_2$ and the corresponding materials represent appealing low energy phase transition materials.
Further thermal effects can be evaluated by adding the contribution of vibrational frequencies to the free energy of each phase.
Similar to charge doping, one can predict additional low strain phase transition materials by adjusting the concentration of the dopant according
to the linear dependency of the concentration with energy.

\begin{table}[h]
\begin{tabular}{ c|ccc|ccc|}
  & \multicolumn{6}{c|}{Critical uniaxial load (\%)} \\
  & \multicolumn{3}{c|}{along a} & \multicolumn{3}{c|}{along b}  \\
  Dopant & nSOC & SOC$^*$ & SOC & nSOC & SOC$^*$ & SOC \\ \hline \hline
 Sc &  +0.50 &  +1.22 & & -0.39 & -0.91 & \\
 Y  &  +0.43 &  +1.12 & & -0.48 & -0.91 & \\
 Ti & -0.97 & {\bf -0.35} & &  +0.89 & {\bf +0.30} & \\
 Zr & -1.59 &  -0.97 & &  +1.75 & +0.98 & \\
 V  & -1.25 &  {\bf -0.55} & &  +1.09 & {\bf +0.46} & \\
 Nb & -1.82 & -1.22 & &  +1.61 & +1.02 & \\
 Cr & -2.13 & -1.65 & & +2.25 & +1.64 & \\ \hline
 Mo & -2.28 & -1.89 & -1.89 &  +2.49 &  +1.88 & +1.92 \\ \hline
 Mn & -0.50 & {\bf +0.56} & {\bf +0.63} & +0.26 & {\bf -0.25} & {\bf -0.28} \\
 Tc & -0.38 & {\bf +0.53} & {\bf +0.90} & +0.24 & {\bf -0.34} & {\bf -0.48} \\
 Fe & +0.51 & +2.39 & & -0.23 & -0.94 & \\
 Ru & +1.84 & +3.23 & & -0.82 & -1.33 & \\
 Co & +2.70 & +4.00 & & -1.49 & -2.30 & \\
 Rh & +4.71 & n/a & & -2.06 & -2.94 & \\
 Ni & +2.04 & +3.18 & & -1.11 & -1.58 & \\
 Pd & +3.78 & n/a & & -2.04 & -2.70 & \\
 \hline
 \end{tabular}
\caption{Critical uniaxial load (in \%) along a and b-directions required to switch the phase of various doped MoTe$_2$ with (SOC) and without SOC (nSOC).
The column SOC$^*$ corresponds to an approximation of the SOC case, as described in the text and in the Method section.}
\label{table:uni-load}
\end{table}

\begin{figure}[ht]
\includegraphics[width=5in,height=3in]{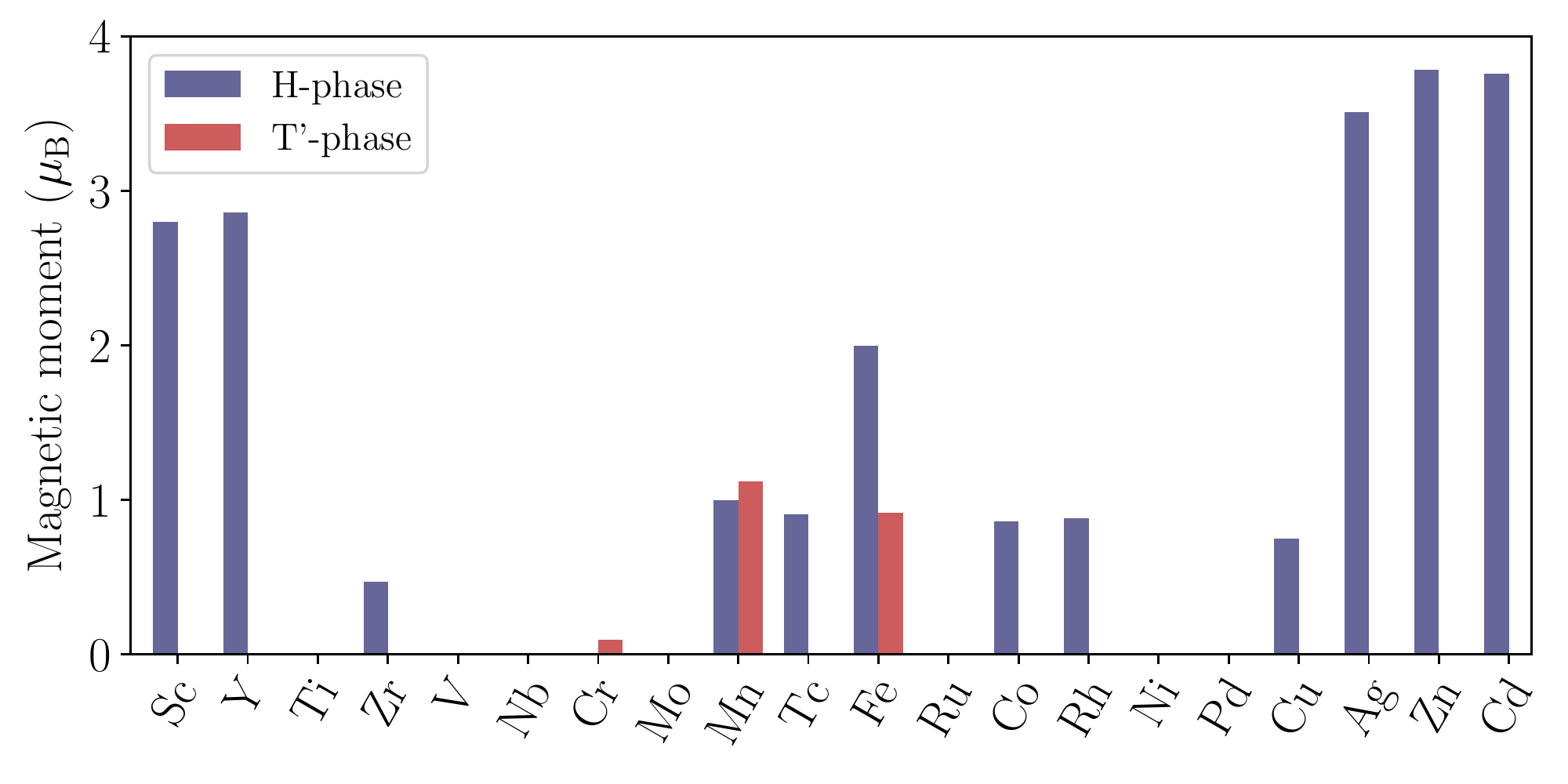}
  \caption{Magnetic moments (in $\mu_B$) for various doped MoTe$_2$ in H and T' phases.}
  \label{fig:magmom}
\end{figure}

{\bf Magnetization.}
The introduction of 3d and 4d TMs in MoTe$_2$ can also induce magnetism in the materials due to their localized d-electrons.
Figure \ref{fig:magmom} shows the magnetic moment of each 6.3 \% doped material in both H and T'-phases.
Ti, V, Nb, Cr, Ru, Ni and Pd doped materials as well as pristine MoTe$_2$ have zero magnetic moments which indicates the non-magnetic nature of these materials.
Mn doped materials shows magnetic moment of approximately 1 Bohr magneton for both phases while Fe doped MoTe$_2$ shows magnetic moments 2 to 1 Bohr
in H-phase and T'-phase, respectively.
The most exciting compounds correspond to Sc, Y, Co, Rh, Cu, Ag, Zn and, Cd doped MoTe$_2$ which show no sign of magnetism in the T'-phase while a significant
magnetic moment in H-phases.
This suggests the possibility not only to tune the semiconductor-to-metal transition but also the magnetic property via electrostatic voltage or strain.
Moreover, it should be possible to design device to magnetically trigger the phase transition in these materials.

\section*{Conclusion} 
To summarize, we systematically investigated the stability of H and T' phases of 3d and 4d TM doped MoTe$_2$ using first principle calculations.
We found that doping can be used to tune the phase transition in MoTe$_2$ and that the mechanism relies on the interplay between LF and CDW.
A small amount of dopant can lower the energy difference between semimetallic T' and semiconducting H phases and we predict the additional
perturbation to trigger the transition via strain and electrostatic voltage.
Compounds such as 6.3 \% Mn or Tc and, 4.2 \% Mn or Ti doped MoTe$_2$ appear as promising candidates for low energy and fast switching materials.
Moreover, we described a simple procedure to predict critical dopant concentration and hence, additional potential dopants.
Interestingly, the introduction of TM dopants significantly lowers the phase transition kinetics and, introduces phase dependent magnetic moment.
This last property can have important implication in nanoelectronics where the structural phase transition could be induced by an electrostatic or
magnetic field\cite{asamitsu1995structural}.
This work provides a guideline to experimentalists to explore the possibility of phase engineering in MoTe$_2$ and can be extended to other 2D materials.

\section*{Methods}

All calculations were based on DFT as implemented in VASP\cite{vasp-1,vasp-2,vasp-3} using plane wave projected-augmented wave (PAW) pseudopotentials\cite{paw}.
The generalized gradient approximation proposed by Perdew, Burke, and Ernzerhof\cite{pbe} was used with an energy cutoff of 350 eV.
The Brillouin zone was sampled on a 2$\times$2$\times$1 Monkhorst-Pack\cite{monk} type mesh to described the 4$\times$4$\times$1 supercell.
To avoid any interaction between layers, we added a vacuum space of 36 \AA.
Lattice parameters and internal coordinates of the ions were relaxed until energy and force of charged supercells reached values of 0.5$\times$10$^{-7}$ and
0.5$\times$10$^{-6}$ eV/f.u. while we used the looser criteria of 0.01 eV/\AA~to converge forces in strained supercells.
All calculations were spin-polarized.
For charge doping, the total number of electrons was changed and ions relaxed within the lattice parameters of the uncharged
ground state structure, corresponding to constant stress case.
Critical excess charges were extracted by linearly fitting the H/T' energy difference as a function of the excess charge around the
zero energy intersection point.
We did not correct for the background charge since this has been demonstrated to cancel when energy differences are considered \cite{santosh}.
We have further validated this in the Supplementary Materials.
H and T' strain surfaces were interpolated with a 2D cubic surface and the intersection contours with cubic curves.
Uniaxial loads were extracted from successive cubic fits of the strain data as described in the Supplementary Materials of Ref.\citep{Reed-14} and
reported section S4 of the Supplementary Materials.

\begin{acknowledgement}
This work was supported by the University Grants Committee of Hong Kong (ECS PolyU-253015).
We thank the Hong Kong Polytechnic University and Purdue University for the computational resources.
\end{acknowledgement}

\begin{suppinfo}
Details on the charge doping procedure, how to convert critical excess charge to voltage, validation cases and energy calculations
as well as the effect of doping content, the kinetics of the phase transition and, uniaxial load procedure and corresponding figures.
\end{suppinfo}


\providecommand{\latin}[1]{#1}
\providecommand*\mcitethebibliography{\thebibliography}
\csname @ifundefined\endcsname{endmcitethebibliography}
  {\let\endmcitethebibliography\endthebibliography}{}

\end{document}